 \documentclass[final,3p,times,twocolumn]{elsarticle}
\usepackage{amssymb}
\usepackage{amsmath}
\usepackage{mathtools}
\usepackage{hyperref}
\usepackage{booktabs}
\usepackage{color}
\usepackage{tabularx}
\usepackage{algorithm}
\usepackage{algpseudocode}

\usepackage{lipsum}

\newcounter{bla}

\journal{Computer Physics Communications}

\begin{document}

\begin{frontmatter}



\title{PairDiagSph: Generalization of the Exact Pairing Diagonalization Program for Spherical Systems}

\author[a,b,c]{Xiao-Yu Liu}
\author[c]{Chong Qi\corref{author}}
\author[d]{Xin Guan}
\author[a]{Zhong Liu}

\cortext[author]{Corresponding author.\\\textit{E-mail address:} chongq@kth.se}
\address[a]{Institute of Modern Physics, Chinese Academy of Sciences, Lanzhou 730000, China}
\address[b]{University of Chinese Academy of Sciences, Beijing 100049, China}
\address[c]{Department of Physics, Royal Institute of Technology, Stockholm 104 05, Sweden}
\address[d]{Department of Physics, Liaoning Normal University, Dalian 116029, China}

\begin{abstract}

We present an efficient program for the exact diagonalization solution of the pairing Hamiltonian in spherical systems with rotational invariance based on the SU(2) quasi-spin algebra. The basis vectors with quasi-spin symmetry considered are generated by using an iterative algorithm. Then the Hamiltonian matrix constructed on this basis is diagonalized with the Lanczos algorithm. All non-zero matrix elements of the Hamiltonian matrix are evaluated ``on the fly" by the scattering operator and hash search acting on the basis vectors. The OpenMP parallel program thus developed, PairDiagSph, can efficiently calculate the ground-state eigenvalue and eigenvector of general spherical pairing Hamiltonians. Systems with dimension up to 10$^{8}$ can be calculated in few hours on standard desktop computers.

\end{abstract}


\end{frontmatter}

\begin{small}
\noindent
{\bf PROGRAM SUMMARY}\\
{\em Program Title:} PairDiagSph.\\
{\em Licensing provisions:} CC by NC 3.0.\\
{\em Programming language:} Fortran 95.\\
{\em Nature of problem:} The exact diagonalization of spherical pairing Hamiltonian can be achieved in the quasi-spin space.\\
{\em Solution method:} The program generates the basis vectors via the adjacency excitation algorithm, and diagonalizes the spherical pairing Hamiltonian by the Lanczos + QR algorithm.\\
{\em Restrictions:} The total number of spherical must be less than 64; The maximal dimension that can be handled is restricted by the local RAM capacity.\\
\end{small}

\section{Introduction}

In a recent paper~\cite{liuu2020}, we developed an efficient Fock-space diagonalization program, PairDiag, for solving the general pairing Hamiltonian in the deformed system with time-reversal invariance. In the program, the basis vector of the Slater determinant is represented by a binary word, where each bit of the word is associated to a pair of doubly-degenerate orbitals and the value of the bit is set to 1 (or 0) when the corresponding paired orbital is fully occupied (or empty). In such a representation, all binary-based vectors are generated in ascending order by the `01' inversion algorithm. The hash search algorithm acting on the basis for directly locating all non-zero Hamiltonian matrix elements improves greatly the efficiency of the Lanczos~\cite{lanc1950} diagonalization process.

Many nuclei near closed shells show behaviors of spherical symmetry. In these rotationally invariant systems, the single-particle levels from the nuclear shell model can have degeneracy higher than 2. Based on the existence of quasi-spin symmetry~\cite{raca1943}, the subset of degenerate levels in paired system can be packaged as a whole with its state labeled by the partial quasi-spin. The spherical pairing Hamiltonian matrix constructed on such the quasi-spin basis will have a dimension much lower than the fermionic Fock space. The exact pairing solution based on the SU(2) quasi-spin algebra is presented in Refs.~\cite{ker11961, voly2001}. However, in actual diagonalization calculations, the generation of quasi-spin basis vectors was usually achieved via the trial and error approach which is time-consuming, and the time complexity of the matrix diagonalization was also high. Therefore, our earlier studies~\cite{xuuu2013, chan2015} were very much limited to small spaces with dimension less than $10^5$.

In this work, we present an extension of the exact pairing solver from PairDiag~\cite{liuu2020} for deformed systems to a new program, PairDiagSph, for spherical systems based on the SU(2) quasi-spin algebra. In the program PairDiagSph, we continue the efficient diagonalization method of using the search algorithm to directly locate all non-zero Hamiltonian matrix elements, while extend both the vector generation and hash search algorithms to the quasi-spin basis system. With the OpenMP parallel~\cite{cham2008} PairDiagSph module, the calculation of spherical pairing Hamiltonian for systems with dimension up to 10$^{8}$ can be completed within hours on standard desktop computers. The developed adjacency excitation algorithm, as a general solution to the balls-into-boxes problem, can also be transplanted to related applications.

\section{The General Pairing Hamiltonian and Quasi-spin Algebras}

The general pairing Hamiltonian in deformed systems with time-reversal invariance is given by 
\begin{equation}
\label{hamiltonian1}
\hat{H} = \sum_{i}\epsilon_{i}(a_{i}^{\dagger}a_{i}^{}+a_{\bar{i}}^{\dagger}a_{\bar{i}}^{})+\sum_{ii'}G_{ii'}a_{i}^{\dagger}a_{\bar{i}}^{\dagger}a_{\bar{i'}}^{}a_{i'}^{}
\end{equation}
\noindent where ${i}$ and $\bar{i}$ are a pair of degenerate time-reversed orbitals, $\epsilon_{i}$ and $G_{ii'}$ are the orbital-dependent single-particle energies and pairing interaction strength, $a_{i}^{\dagger}$ and $a_{i}^{}$ is the particle creation and annihilation operator, respectively. In Ref.~\cite{liuu2020}, we developed a Fock-space diagonalization program to solve this Hamiltonian. For spherical systems with rotational invariance, the pairing Hamiltonian can be generalized~\cite{ker11961, voly2001} as
\begin{equation}
\label{hamiltonian2}
\hat{H} = \sum_{jm}\epsilon_{j}a_{jm}^{\dagger}a_{jm}^{}+\frac{1}{4}\sum_{jj'}G_{jj'}\sum_{mm'}a_{jm}^{\dagger}\tilde{a}_{jm}^{\dagger}\tilde{a}_{j'm'}^{}a_{j'm'}^{}
\end{equation}
\noindent where $\tilde{a}_{jm} = (-1)^{j-m}a_{j-m}$, $j$ represents a single-particle orbitals with angular momentum $j$ and degeneracy $2j+1$, $\epsilon_{j}$ are the single-particle energies for the involved orbitals, $G_{jj'}$ are the pairing interaction strength between the orbital $j$ and $j'$. In the present work, we will solve this Hamiltonian based on quasi-spin~\cite{raca1943} algebras.

The Hamiltonian in Eq.~(\ref{hamiltonian2}) can be rewritten as
\begin{equation}
\label{hamiltonian3}
\hat{H} = \sum_{j}\left(\epsilon_{j}(2L_{j}^{z}+\Omega_{j})+G_{jj}L_{j}^{+}L_{j}^{-}\right)+\sum_{j\neq j'}G_{jj'}L_{j}^{+}L_{j'}^{-}
\end{equation}
\noindent by introducing the quasi-spin $L_{j}^{\pm}$ and $L_{j}^{z}$ operators~\cite{voly2001, Talm1993, kerm1961, ring1980} for each single $j$ shell as
\begin{equation}
\label{quasispin}
\begin{split}
& L_{j}^{+} = \frac{1}{2}\sum_{m}a_{jm}^{\dagger}\tilde{a}_{jm}^{\dagger} \\
& L_{j}^{-} = (L_{j}^{+})^{\dagger} = \frac{1}{2}\sum_{m}\tilde{a}_{jm}^{}a_{jm}^{} \\
& L_{j}^{z} = \frac{1}{2}\sum_{m}(a_{jm}^{\dagger}a_{jm}^{}-\frac{1}{2}) = \frac{1}{2}(N_{j}-\Omega_{j})\\
\end{split}
\end{equation}
\noindent where $N_{j}$ is the particle number and $\Omega_{j} = (2j\!+1)/2$ is the pair degeneracy. Based on the following commutation relations
\begin{equation}
\label{commutation}
\begin{split}
& [L_{j\;}^{+}, L_{j'}^{-}] = 2\delta_{jj'}L_{j}^{z} \\
& [L_{j\;}^{z}, L_{j'}^{\pm}] = \pm\delta_{jj'}L_{j}^{\pm} \\
\end{split}
\end{equation}
\noindent we see that these quasi-spin operators form an SU(2) algebra with $L_{j}^{\pm}$ and $L_{j}^{z}$ corresponding to the raising/lowering and the $z$-component angular momentum operators, respectively. The square of quasi-spin with eigenvalue $L_{j}(L_{j}\!+\!1)$ can be defined as $L_{j}^{2}=L_{j}^{+}L_{j}^{-}+(L_{j}^{z})^{2}-L_{j}^{z}$. The maximum value of $L_{j}$ is $\Omega_{j}/2$ for the fully paired orbital. $L_{j}$ could also take lower values as $(\Omega_{j}\!-\!s_{j})/2$ due to the Pauli blocking from the $s_{j}$ unpaired particles in the $j$-th orbital, and $s_{j}$ is usually called the seniority number~\cite{raca1943} which is conserved under the pairing Hamiltonian. With the quasi-spin symmetry, we can use $L_{j}$ and $L_{j}^{z}$ to label a state of a single $j$ shell as $|L_{j},L_{j}^{z}\rangle$.

In a system of $m$ shells with $N\!=\!\sum_{j}^{m}\!N_{j}$ particles and fixed seniority $S\!=\!\sum_{j}^{m}\!s_{j}$, there will be $n\!=\!(N\!-S)/2$ particle pairs formed, and we can define for the $j$-th shell the pair number $n_{j}$ and the effective pair degeneracy $\omega_{j}$ as
\begin{equation}
\label{news}
\begin{split}
& n_{j} = \frac{(N_{j}-s_{j})}{2} \\
& \omega_{j} = \Omega_{j}-s_{j} \\
\end{split}
\end{equation}
\noindent For such a orbital with fixed $s_{j}$, it is more convenient to use only the pair number $n_{j}$ to label its state as $|n_{j}\rangle$. We can construct quasi-spin basis vectors for the system as $|n_{1},n_{2},\cdots\!,n_{m}\rangle$ exhausting all possible solutions of $n\!=\!\sum_{j}^{m}\!n_{j}$ with constrains $0\!\le\!n_{j}\!\le\!\omega_{j}$. Then,  the Hamiltonian matrix in Eq.~(\ref{hamiltonian3}) can be constructed on this basis by using the following relations
\begin{equation}
\label{relations}
\begin{split}
& L_{j}|n_{j}\rangle = \frac{\Omega_{j}-s_{j}}{2}|n_{j}\rangle = \frac{\omega_{j}}{2}|n_{j}\rangle \\
& L_{j}^{z}|n_{j}\rangle = \frac{N_{j}-\Omega_{j}}{2}|n_{j}\rangle = (n_{j}-\frac{\omega_{j}}{2})|n_{j}\rangle \\
& L_{j}^{\pm}|n_{j}\rangle = \sqrt{(L_{j}\mp L_{j}^{z})(L_{j}\pm L_{j}^{z}+1)}\,|n_{j}\!\pm\!1\rangle \\
\end{split}
\end{equation}
\noindent Diagonal elements from the first term of the Hamiltonian become
\begin{equation}
\label{diagonal}
\begin{split}
& \langle \cdots,n_{j},\cdots|\hat{H}|\cdots,n_{j},\cdots \rangle = \qquad\qquad\qquad\quad \\
& \ \ \, \sum_{j}\left(2\epsilon_{j}n_{j} + G_{jj}n_{j}(\omega_{j}-n_{j}+1)\right) + \sum_{j}\epsilon_{j}s_{j} \\
\end{split}
\end{equation}
\noindent Non-diagonal elements described by the second term which scatters a pair between the orbital $j$ and $j'$ are
\begin{equation}
\label{nondiagonal}
\begin{split}
& \langle\cdots\!,n_{j}\!+\!1,\!\cdots\!,n_{j'},\!\cdots|\hat{H}|\cdots\!,n_{j},\!\cdots\!,n_{j'}\!+\!1,\!\cdots\rangle = \\
& \quad G_{jj'}\sqrt{(n_{j}+1)(\omega_{j}-n_{j})}\sqrt{(n_{j'}+1)(\omega_{j'}-n_{j'})} \\
\end{split}
\end{equation}

From Eqs.~(\ref{diagonal}) and (\ref{nondiagonal}) one can see that a system under the spherical pairing Hamiltonian with fixed seniority can be viewed as the sum of two subsystems: One is a non-interactive subsystem of $S$ unpaired particles which will contribute a term $\sum_{j}\!\epsilon_{j}s_{j}$ equally to all diagonal elements; And the other one is a seniority-zero subsystem of $N\!-\!S$ paired particles distributed in orbitals with pair degeneracy reduced to $\omega_{j}\!=\!\Omega_{j}\!-\!s_{j}$.

\section{Principles of the Method}

In the present PairDiagSph program, we solve the spherical pairing Hamiltonian in Eq.~(\ref{hamiltonian2}) or Eq.~(\ref{hamiltonian3}) for  a given seniority via the quasi-spin-space diagonalization to get the ground-state eigenvalue and the corresponding eigenvector. In the following content, we will focus on the solution for seniority-zero systems. The method used can be divided into two parts: Firstly generating the seniority-zero quasi-spin basis; Then diagonalizing the Hamiltonian matrix constructed on the basis.

\subsection{Basis Generation}

Let us consider a seniority-zero system of $m$ shells with degeneracy $\omega_{j=1,2,\cdots,m}$, if there are $n$ $(n\!\le\!\sum_{j}^{m}\!\omega_{j})$ particle pairs placed, the basis with quasi-spin symmetry considered should consist of all possible vectors $|n_{1},n_{2},\cdots\!,n_{m}\rangle$ in which $n\!=\!\sum_{j}^{m}\!n_{j}$ and $0\!\le\!n_{j}\!\le\!\omega_{j}$. Each vector can be represented by a binary word in the computer, while $\omega_{j}$ consecutive bits of the word being associated to the shell $j$, with the number of bits occupied by `\texttt{1}' in the segment depending on the corresponding occupation $n_{j}$. For each degenerate orbital, we will place all $n_{j}$ occupied bits from the lowest digit side of the segment to uniquely mark the state out of all other different permutations since the pair number is the only information needed. Following the rules above, a set of binary numbers with $n$ occupied bits distributed in the first $\sum_{j}^{m}\!\omega_{j}$ digits is equivalent to the seniority-zero quasi-spin space for the system. In the case where 3 pairs occupy 3 shells with degeneracy $\omega_{1,2,3}=\{4,2,1\}$, a set of 6 binary numbers from \texttt{0.00.0111} to \texttt{1.11.0000} (in which the decimal points are just for separating different orbitals) can be used to represent the basis from $|3,0,0\rangle$ to $|0,2,1\rangle$.

\begin{algorithm}
\caption{\label{conversion} Adjacency excitation algorithm. $f(i)$ and $l(i)$ correspond to $f_{i}$ and $l_{i}$ in the text, respectively. BTEST(), IBCLR(), and others refer to the Fortran intrinsic bit manipulation functions}
\begin{algorithmic}
\Require integer I$_{in}$
\Ensure integer I$_{out}$
\State I$_{tail}$ = 0
\For{i = 1, $\cdots$, m}
\If {(BTEST(I$_{in}$, $f(i)$))}
\For{j = $f(i)$, $\cdots$, $l(i)$}
\If {(BTEST(I$_{in}$, j))}
\State I$_{in}$ = IBCLR(I$_{in}$, j)
\State I$_{tail}$ = IBSET(ISHFT(I$_{tail}$, 1), 0)
\Else
\State \textbf{exit}
\EndIf
\EndFor
\If {(!BTEST(I$_{in}$, $l(i+1)$))}
\For{j = $f(i+1)$, $\cdots$, $l(i+1)$}
\If {(!BTEST(I$_{in}$, j))}
\State I$_{in}$ = IBSET(I$_{in}$, j)
\State I$_{tail}$ = ISHFT(I$_{tail}$, -1)
\State \textbf{exit}
\EndIf
\EndFor
\State \textbf{exit}
\EndIf
\EndIf
\EndFor
\State I$_{out}$ = I$_{in}$ + I$_{tail}$
\State \textbf{return} I$_{out}$
\end{algorithmic}
\end{algorithm}

For a system of $n$ identical pairs in $m$ given shells, there is no simple formula to calculate the space dimension directly, this number is usually counted after all possible vectors are created. In the present program, an iterative approach developed based on the `01' inversion algorithm~\cite{liuu2020} was used to generate all the binary-based vectors. Every iteration of the approach takes a binary integer in the space as input, and searches from the first shell (which represented by a segment of $\omega_{j}$ consecutive bits) until the 2 adjacent shells with a specific pattern is found, where the lower shell is filled by at least 1 pair while the higher shell is not fully occupied. Then 1 occupied bit in the lower orbital will be moved to the higher orbital, and all bits `\texttt{1}' below this higher will be moved to refill this integer from the lowest digit. After the two steps, a larger integer in the set is obtained which will be the input for the next iteration. Since in every iteration there will be one pair being excited to its adjacent higher orbital, we call this method the adjacency excitation algorithm. If the degeneracy of each orbital is $\omega_{j}\!=\!1$, the described adjacency excitation algorithm will be simplified to `01' inversion algorithm.

Since we have stipulated that the occupation of each shell starts from its lowest digit in the bit segment, the shell must be empty if the corresponding first digit is empty, and the shell will be full only after the last digit is occupied. For a state \texttt{001.1111.00000} in a system where 5 pairs occupy 3 shells with degeneracy $\omega_{1,2,3}=\{5,4,3\}$, the 1st bit in vacancy indicates the 1st shell is empty, the fully occupied 2nd shell can be reflected in the occupied 9th bit, and based on status of the 10th and the 12th bits, we know the 3rd shell is occupied but not fully. In a system with given degeneracy for $m$ shells, the position of the first and the last digits in the segment corresponding to the $i$-th orbital can be calculated as $f_{i}\!=\!1\!+\!\sum_{j}^{i-1}\!\omega_{j}$ and $l_{i}\!=\!\sum_{j}^{i}\!\omega_{j}$. With these definition, a pseudocode of the adjacency excitation algorithm based on Fortran bit operations is shown in Algorithm~\ref{conversion}. In the practical calculation for a space with dimension $n$, the minimum and the maximum vectors in the space representing the start and the end of the iteration must be specified in advance, and the remaining vectors can be generated from the minimum within $n\!-\!1$ times of iteration. So, the time complexity the algorithm over the entire space can be roughly estimated as a linear order $O(n)$.

For the previous example with the minimum \texttt{0.00.0111} and the maximum \texttt{1.11.0000}, the iteration should start at \texttt{0.00.0111} and end when the output reaches \texttt{1.11.0000}. In the first iteration, a pair in the first shell needs to be excited to the second to form the output \texttt{0.01.0011}. In the same way \texttt{0.11.0001} is the second output. For the input \texttt{0.11.0001} in which the second shell is fully occupied, \texttt{1.01.0001} is obtained after moving a pair from the second shell to the third, then the remaining pairs below the third shell need to be de-excited to the lowermost to get the output \texttt{1.00.0011}. Iteratively, \texttt{1.01.0001} and \texttt{1.11.0000} will be created in order, and then the iteration should be terminated as the \texttt{1.11.0000} reaches the maximum. With five iterations, all six integers obtained are summarized in Table~\ref{data}, in which the indexes are assigned in the order of generation. In PairDiagSph program, a 64-bit integer is used to represent a basis vector and all the generated integers are stored sequentially in an 1D array. The capacity $\sum_{j}^{m}\!\omega_{j}$ of the system should be less than 64 due to the sign bit. Since the vector array is strictly in ascending order and organized by special combination rules, the index $i$ of any element can be calculated via search algorithm from its binary value $|i\rangle$ which represent a specific quasi-spin wave function $|n_{1},n_{2},\cdots\!,n_{m}\rangle$.

\begin{table}
\small
\begin{center}
\setlength{\tabcolsep}{2.9mm}{
\caption{\label{data} Index and binary values of all integers and the corresponding wave function $|n_{1},n_{2},n_{3}\rangle$ in the space of 3 pairs in 3 degenerate shells of $\omega_{1,2,3}=\{4,2,1\}$. The decimal points in the binary values are just for separating different orbitals. The decimal values shown displays the ascending order.}
\begin{tabular}{cccc}
\hline
Index  &  Binary value  &  $|n_{1},n_{2},n_{3}\rangle$ &  Decimal value \\
\hline
1      &  0.00.0111     &  $|3,0,0\rangle$             &  007           \\
2      &  0.01.0011     &  $|2,1,0\rangle$             &  019           \\
3      &  0.11.0001     &  $|1,2,0\rangle$             &  049           \\
4      &  1.00.0011     &  $|2,0,1\rangle$             &  067           \\
5      &  1.01.0001     &  $|1,1,1\rangle$             &  081           \\
6      &  1.11.0000     &  $|0,2,1\rangle$             &  112           \\
\hline
\end{tabular}}
\end{center}
\end{table}

\subsection{Vector Search}

In PairDiagSph program, an efficient hash search algorithm with the time complexity $O(1)$ is built to locate the index $i$ of an element $|i\rangle$ in the generated basis array. For all basis vectors in a $p$-pairs system with given degeneracy, We define $N^{p}_{d}$ as the minimum number of iterations required to generate a binary-based vector with the $d$-th ($\sum_{j}\!\omega_{j}\!\ge\!d\!\ge\!p$) digit occupied from the minimum vector. For the first 4 items in Table~\ref{data}, we can get in that 3-pairs system $N^{3}_{3}=0$, $N^{3}_{5}=1$, $N^{3}_{6}=2$, and $N^{3}_{7}=3$. Except for $N^{p}_{p}\!=\!0$, the value of $N^{p}_{d}$ with $d\!>\!p$ is degeneracy dependent. With the definition of $N^{p}_{d}$, the index of any vector in an untruncated space generated by the adjacency excitation algorithm can be expressed as the sum of a series $N^{p}_{d}$ with different $p$ and $d$. Let us take a vector \texttt{0011.01111.001.0001} as an example, we first need $N^{8}_{14}$ steps of iteration to generate the vector \texttt{0011.00000.011.1111} from the minimum \texttt{0000.00000.011.1111}, then another $N^{6}_{11}$ steps are needed to convert the vector \texttt{0011.00000.011.1111} to \texttt{0011.01111.000.0011}, finally, vector \texttt{0011.01111.001.0001} will be obtained after $N^{2}_{5}$ times of iteration based on \texttt{0011.01111.000.0011}. So its index can be counted as $i=1\!+\!N^{8}_{14}\!+\!N^{6}_{11}\!+\!N^{2}_{5}$. For an arbitrary vector $|i\rangle$ in a system, we can define for the $i$-th shell the $p_{i}=\sum_{j}^{i}\!n_{j}$ and $d_{i}=n_{i}\!+\!\sum_{j}^{i-1}\!\omega_{j}$, then the hash function $i\!=\!f(|i\rangle)$ for the search can be written as
\begin{equation}
\label{hash}
f(|i\rangle) = 1+\sum_{j}^{m}(1-\delta_{n_{j},0})N^{p_{j}}_{d_{j}}
\end{equation}

In general, the hash search for a $n$-pairs system requires all possible coefficients $N^{p}_{d}$ with $p\!\le\!n$ and $d\!\le\!\sum_{j}\!\omega_{j}$, and there is no simple formula to calculate them directly. One feasible way to get these coefficients is to solve them backwards in the linear equations of all the hash functions in Eq.~(\ref{hash}) with known indexes. In the basis system, the $i$-th vector is obtained by performing the adjacency excitation operation on the $(i\!-\!1)$-th vector, and only the excitation operation can introduce a new coefficient in the corresponding $i$-th hash equation compared with the $(i\!-\!1)$-th equation. The first equation with index 1 contains no coefficient and each subsequent equation will introduce at most one unknown new coefficient, this means these linear equations can be easily solved in order from the second one till the last, and it is also undoubtedly correct when we use these coefficients to calculate the indexes back. Still taking the vectors in Table~\ref{data} as an example, each vector with its index corresponds to a linear hash equation, and all these equations listed in Table~\ref{equa} can be solved easily in order. In PairDiagSph program, all the required coefficients are calculated during the generation of the basis vectors, and then these results are stored in a 2D array which will be used as a table in the later hash search. 

\begin{table}
\small
\begin{center}
\setlength{\tabcolsep}{3.4mm}{
\caption{\label{equa} In the space of 3 pairs in 3 degenerate orbitals of $\omega_{1,2,3}=\{4,2,1\}$, all the binary-based vectors, hash equations, and the corresponding solutions.}
\begin{tabular}{ccll}
\hline
Index &  Binary value &  Equation                     &  Solution       \\
\hline
1     &  0.00.0111    &  1=1                          &                 \\
2     &  0.01.0011    &  2=1+$N^{3}_{5}$              &  $N^{3}_{5}=1$  \\
3     &  0.11.0001    &  3=1+$N^{3}_{6}$              &  $N^{3}_{6}=2$  \\
4     &  1.00.0011    &  4=1+$N^{3}_{7}$              &  $N^{3}_{7}=3$  \\
5     &  1.01.0001    &  5=1+$N^{3}_{7}$+$N^{2}_{5}$  &  $N^{2}_{5}=1$  \\
6     &  1.11.0000    &  6=1+$N^{3}_{7}$+$N^{2}_{6}$  &  $N^{2}_{6}=2$  \\
\hline
\end{tabular}}
\end{center}
\end{table}

\subsection{Matrix Construction and Diagonalization}

With the basis generated and the search algorithm provided, we can now construct the pairing Hamiltonian matrix in an efficient way by evaluating all non-zero matrix elements directly. The diagonal elements in the Hamiltonian matrix are usually non-zero and the value of $H_{i,i}$ can be calculated from Eq.~(\ref{diagonal}). Of all non-diagonal elements $H_{i,j}$, only a small part of them are non-zero. For a vector $|i\rangle$ with index $i$ in a system of $m$ shells, if we mark one shell of $n_{P}\!>\!0$ as $P$ and another shell of $n_{V}\!<\!\omega_{V}$ as $V$, then ``scatter" 1 pair from shell $P$ to $V$ to form a new vector $|j\rangle\!=\!L^{+}_{V}L^{-}_{P}|i\rangle$, the matrix element $H_{i,j}\!=\!\langle j|\hat{H}|i\rangle$ described in Eq.~(\ref{nondiagonal}) will be non-zero (if $G_{VP}\!\neq\!0$). The position of this element $(i, j)$ in matrix can be obtained by searching the index $j$ of vector $|j\rangle$. Combining the different $P$ and $V$ in $|i\rangle$, the maximum number of such $|j\rangle$ and also the non-zero $H_{i, j}$ is $m(m\!-\!1)$. Still using the previous example in Table~\ref{data} with assigning single-particle energies $\epsilon_{1,2,3}\!=\!\{1,2,3\}$ and the constant $G_{i,j}\!=\!-0.2$ as the overall pairing interaction strength. The Hamiltonian can be expressed as a $6\!\times\!6$ real symmetric matrix. For the first row, the diagonal element is $H_{1, 1}\!=\!\langle 3,0,0|\hat{H}|3,0,0\rangle\!=\!4.8$. The 2 non-diagonal non-zero elements are $H_{1,2}$ from $|2,1,0\rangle\!=\!L^{+}_{2}L^{-}_{1}|3,0,0\rangle$ with value $\langle 2,1,0|\hat{H}|3,0,0\rangle\!=\!-0.2\!\sqrt{12}$, and $H_{1,4}$ from $|2,0,1\rangle\!=\!L^{+}_{3}L^{-}_{1}|3,0,0\rangle$ with value $\langle 2,0,1|\hat{H}|3,0,0\rangle\!=\!-0.2\!\sqrt{6}$. The rest rows of the matrix can also be constructed this way.

For diagonalizing the obtained Hamiltonian matrix, we use the same Lanczos~\cite{Wuxx2000}+QR~\cite{saad2011} method as in the PairDiag program~\cite{liuu2020}. All the non-zero matrix elements which are mainly used for matrix-vector multiplication in Lanczos iterations are evaluated directly ``on the fly" to reduce time and space complexity of the calculation. Depending on the user's choice, the PairDiagSph program can return the ground state eigenvalue and eigenvector after \texttt{Lanc\_\,Limit} times of iteration, or perform the restart Lanczos in which the calculation will be restarted by the ground-state Ritz vector with the convergence condition $|\,\beta_{i}/\alpha_{i}| \le$ \texttt{Lanc\_\,Error}~\cite{liuu2020}. In the program, the adjustable parameters, \texttt{Lanc\_\,Limit}, is the subspace dimension of the Lanczos iteration, and the larger \texttt{Lanc\_\,Limit} will lead to higher quality results by the cost of more RAM memory. For the calculation with dimension $N$ and \texttt{Lanc\_\,Limit} = $R$, the memory needed to store the basis and Lanczos/Ritz vectors is about $8N(R\!+\!2)\!\times\!10^{-9}$GB in total, which means at least 41.6GB of memory is required for $N=10^{8}$ and \texttt{Lanc\_\,Limit} $=50$. Users need to adjust \texttt{Lanc\_\,Limit} to fit their local RAM conditions, and a larger value is recommended whenever possible. Another parameters, \texttt{Lanc\_\,Error}, is the convergence threshold for the restart Lanczos, and the smaller \texttt{Lanc\_\,Error} will lead to higher quality results by the cost of more times of restart. The predefined \texttt{Lanc\_\,Error} = $1\!\times\!10^{-5}$ in the program meets general accuracy requirements. More details about the diagonalization process and computational performance is presented in Ref.~\cite{liuu2020}.

\section{Description of the Code}

The PairDiagSph code is written in Fortran 95 and packaged in a Fortran module called \texttt{PairDiagSph}. The use of the module, as an example shown in~\ref{example}, requires the following steps.

\subsection{Step 0. Declare a Variable of the Type Diag\_\,Par} 
 
The \texttt{PairDiagSph} module needs to be loaded into the local program before use. After the loading, a predefined derived data type, \texttt{Diag\_\,Par}, will be available which contains 9 components:

\begin{itemize}
\setlength{\abovedisplayskip}{0pt}
\setlength{\itemsep}{0pt}
\setlength{\parsep}{0pt}
\setlength{\parskip}{0pt}
\item \texttt{Shell}: \texttt{Integer(kind=8)}.
\item \texttt{Pairs}: \texttt{Integer(kind=8)}.
\item \texttt{Omega}: \texttt{Integer(kind=8),dimension(63)}.
\item \texttt{Senio}: \texttt{Integer(kind=8),dimension(63)}.
\item \texttt{SPE}: \texttt{Real(kind=8),dimension(63)}.
\item \texttt{P\_\,F}: \texttt{Real(kind=8),dimension(63, 63)}.
\item \texttt{Energy\_\,Ground}: \texttt{Real(kind=8)}.
\item \texttt{Monopole\_\,Min}: \texttt{Real(kind=8)}. 
\item \texttt{N\_\,Occup}: \texttt{Real(kind=8),dimension(63)}.
\end{itemize}

Users need to declare a variable of the type \texttt{Diag\_\,Par} in their program, and this variable, a mandatory parameter of the calculation, will be used to pass parameters (with the first 6 components) and receive results (with the last 3 components).

\subsection{Step 1. Initialize the Input Part}

The first 6 components in the \texttt{Diag\_\,Par} variable which represent the inputs for the calculation must be explicitly initialized by users.

\begin{itemize}
\setlength{\abovedisplayskip}{0pt}
\setlength{\itemsep}{0pt}
\setlength{\parsep}{0pt}
\setlength{\parskip}{0pt}
\item \texttt{Shell}: The total number of shells, $m$.
\item \texttt{Pairs}: The total number of pairs, $n$.
\item \texttt{Omega}: 1D array for the degeneracy, $\Omega_{j}$.
\item \texttt{Senio}: 1D array for the seniority, $s_{j}$. 
\item \texttt{SPE}: 1D array for the single-particle energy, $\epsilon_{j}$.
\item \texttt{P\_\,F}: 2D array for the pairing interaction strength, $G_{jj'}$.
\end{itemize}

The value of \texttt{Shell} ($m$) should be no more than 63, and the first $m$ terms of the 1D array (\texttt{Omega}, \texttt{Senio}, and \texttt{SPE}) and the first $m\!\times\!m$ part of the 2D array \texttt{P\_\,F} will be used to construct the basis and the Hamiltonian matrix. Users also need to ensure that $0\!\le\!\omega_{j}\!=\!\Omega_{j}-\!s_{j}$ for each shell, and $n\!\le\!\sum_{j}^{m}\!\omega_{j}\!\le\!63$ for the system. The pairing interaction matrix $G$ should be initialized in a real symmetric manner. The total particle number of the system is $N\!=\!2n+\!\sum_{j}^{m}\!s_{j}$. 

There are three parameters that can be optionally adjusted in the source code PairDiagSph.f90.

\begin{itemize}
\setlength{\itemsep}{0pt}
\setlength{\parsep}{0pt}
\setlength{\parskip}{0pt}
\item \texttt{Lanc\_\,Limit}: The size of the Lanczos iteration subspace, the default value is 50 and the recommended range is between 10 and 50 for the ground state. 
\item \texttt{Lanc\_\,Error}: The convergence threshold for the restart Lanczos calculation, the default value $1\!\times\!10^{-5}$ meets general accuracy requirements.
\item \texttt{Print\_\,Mode}: Only when the value is 0 (default), the program will print information on the terminal.
\end{itemize}

\subsection{Step 2. Call the Subroutine}

There is only one public subroutine that can be called in the PairDiagSph module.

\begin{itemize}
\setlength{\itemsep}{0pt}
\setlength{\parsep}{0pt}
\setlength{\parskip}{0pt}
\item \texttt{Diag\_\,Solver(Diag\_\,Par, [Mode])}: The public subroutine calculates the pairing Hamiltonian in Eq.~(\ref{hamiltonian2}) for the system described by the input part of the \texttt{Diag\_\,Par} variable. The optional parameter, \texttt{Mode}, will affect the process of Lanczos, \texttt{Mode = 0} (default) corresponds to the restart Lanczos, and \texttt{Mode = 1} corresponds to the Lanczos without restart.
\end{itemize}

\subsection{Step 3. Analyze the Output Part}

The results of the calculation is stored in the last 3 components of the  \texttt{Diag\_\,Par} variable. The occupation numbers are saved instead of the full eigenvectors to save space.

\begin{itemize}
\setlength{\itemsep}{0pt}
\setlength{\parsep}{0pt}
\setlength{\parskip}{0pt}
\item \texttt{Energy\_\,Ground}: The eigenvalue of ground state corresponding to the $\langle\phi_{g.s.}|\hat{H}|\phi_{g.s.}\rangle$.
\item \texttt{Monopole\_\,Min}: The minimum diagonal element of the pairing Hamiltonian matrix (this value is not the HF energy).
\item \texttt{N\_\,Occup}: 1D array for the occupation numbers corresponding to $(2\langle\phi_{g.s.}|n_{j}|\phi_{g.s.}\rangle\!+\!s_{j})$.
\end{itemize}

A simple program example for the spherical pairing Hamiltonian using the PairDiagSph module can be found in~\ref{example}. Users can  modify the program according to their own requirements. A brief description of the variables and subroutines in the module is presented in ~\ref{brief}.

\subsection{Parallelization and Compilation}

The parallelization of the program is done for the matrix construction and diagonalization parts. In the present program, only OpenMP~\cite{cham2008} parallelism has been implemented. The code runs in OpenMP parallel mode by default after being compiled with the \texttt{-fopenmp} option in the provided Makefile. The number of parallel threads is not set by the code, so the user can set the environment variable \texttt{OMP\_\,NUM\_\,THREADS} to the desired number. The PairDiagSph program has been tested under both the ifort and gfortran compilers in the Linux system, and we recommend the ifort compiler due to the higher efficiency and stability shown.

\section{Discussion}

We now briefly discuss the performance of the PairDiagSph program. The reference machine is a desktop computer with an Intel Core i7-7700K 4.2GHz$\times$8 CPU and a total of 47GB memory. The compiler used is the Intel Fortran compiler (ifort version 19.0.0.117) under the Ubuntu 16.04 system. We will show below calculations in the model space consisting of 16 spherical orbitals between the magic numbers 20 and 126, including $1f_{7/2}$, $2p_{3/2}$, $1f_{5/2}$, $2p_{1/2}$, $1g_{9/2}$, $1g_{7/2}$, $2d_{5/2}$, $2d_{3/2}$, $3s_{1/2}$, $1h_{11/2}$, $1h_{9/2}$, $2f_{7/2}$, $2f_{5/2}$, $3p_{3/2}$, $3p_{1/2}$, and $1i_{13/2}$. For simplicity, the single-particle energies of these orbitals take integers from 1 to 16, and the constant pairing interaction strength $G_{ij}=G$ is used. 

\subsection{Dimension of the System}

\begin{figure}
\begin{center}
\setlength{\abovecaptionskip}{0.0cm}
\includegraphics[width=0.44\textwidth]{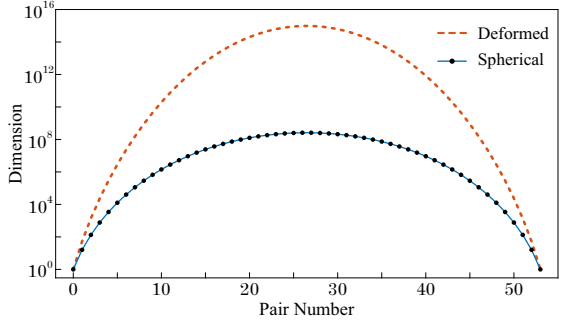}
\caption{\label{dimen} Dimensions of the seniority-zero system as a function of the pair number for spherical (blue solid line with dot symbol) and deformed (red dashed line) systems in the model space consisting of 16 orbitals between the magic numbers 20 and 126.}
\end{center}
\end{figure}

In the defined model space with 16 single $j$ shells and the degeneracy of each orbital as $(2j+1)/2$, the total pair capacity of the space is 53. Fig.~\ref{dimen} shows the relationship between the dimension and the number of pairs in the space. For comparison, we also plotted the dimension of the doubly-degenerate deformed systems in the same model space. Within the quasi-spin symmetry, we can treat all $\omega_{j}$ paired orbitals in a single $j$ shell identically and ignore the different permutations inside the shell. This is the reason why the dimension of the quasi-spin space is greatly reduced (upto six orders of magnitude for the given example). Even at half-filling with 26 pairs, the dimension of the system is only about $2.5\!\times\!10^{8}$. Under the framework of general pairing Hamiltonian in Eq.~(\ref{hamiltonian1}), the dimension of the system in the fermionic Fock space can be calculated by the binomial coefficient $C_{m}^{n}$. In that case, the dimension  at half-filling is as large as $C_{53}^{26}\approx9.7\!\times\!10^{14}$ which is far beyond the current computing power. 

\subsection{Comparison with Other Programs}

\begin{table}
\small
\begin{center}
\setlength{\tabcolsep}{1.6mm}{
\caption{\label{lapack} In the system where 5 pairs in the 16 orbitals, numerical comparisons between PairDiagSph and Lapack. $G$ are the constant pairing interaction strength, $E_{PairDiagS\!ph}$ and $E_{Lapack}$ are the ground-state eigenvalues, $\Delta_{vector}$ is defined as $\sum|V^{2}_{PairDiagS\!ph}(i)-V^{2}_{Lapack}(i)|$, where $V_{PairDiagS\!ph}$ and $V_{Lapack}$ are the calculated ground-state eigenvectors.}
\begin{tabular}{clll}
\hline
$G$            &  $E_{PairDiagS\!ph}$         &  $E_{Lapack}$                &  $\Delta_{vector}$    \\
\hline
\texttt{-}0.2  &  \texttt{+}04.884881026085   &  \texttt{+}04.884881026084   &  3$\times$10$^{-14}$  \\
\texttt{-}0.4  &  \texttt{-}27.750623666024   &  \texttt{-}27.750623666024   &  1$\times$10$^{-12}$  \\
\texttt{-}0.6  &  \texttt{-}70.518391792817   &  \texttt{-}70.518391792817   &  1$\times$10$^{-14}$  \\
\hline
\end{tabular}}
\end{center}
\end{table}

Below we show the numerical performance of PairDiagSph module by comparing with other programs. First, we use the diagonalization results from the Lapack package as a reference. In the system where 5 pairs distributed in the 16 shells, we compared the results of ground state between the two packages with different pairing interaction strengths (for the calculations with PairDiagSph, \texttt{Lanc\_\,Limit} was set to 50). In Table~\ref{lapack}, we present the ground-state eigenvalues and the difference of eigenvectors from the two packages under different constant pairing interaction strengths $G$ (for eigenvectors in PairDiagSph, the user can access \texttt{Q\_\,Matrix} in the subroutine \texttt{Result\_\,Output()} described in~\ref{brief}). Compared with Lapack, the negligible difference between the results indicates that the diagonalization calculation of PairDiagSph is reliable.

\begin{table}
\small
\begin{center}
\setlength{\tabcolsep}{7.5mm}{
\caption{\label{hspa} Numerical comparisons between PairDiagSph and Richardson method. $E_{PairDiagS\!ph}$ and $E_{Richardson}$ are the ground-state eigenvalues, $N_{pair}$ varying from 1 to 6 is the pair number in the predefined space. In calculations the constant pairing interaction strength $G=0.1$ is used.}
\begin{tabular}{cll}
\hline
$N_{pair}$  &  $E_{PairDiagS\!ph}$  &  $E_{Richardson}$  \\
\hline
1           &  1.3198280            &  1.31983           \\
2           &  2.9451548            &  2.94515           \\
3           &  4.8969717            &  4.89697           \\
4           &  7.2131106            &  7.21311           \\
5           &  10.825833            &  10.8258           \\
6           &  14.896384            &  14.8964           \\
\hline
\end{tabular}}
\end{center}
\end{table}

For the standard pairing problem (the Hamiltonian in Eq.~(\ref{hamiltonian1}) with constant interaction strengths $G$), the eigenvalues can also be obtained by the Richardson algebraic approach~\cite{rich1963, rich1964}. In Ref~\cite{guan2020}, a new Numerical algorithm was established for the exact solution of the standard pairing Hamiltonian based on the Richardson-Gaudin method~\cite{gaud1976, pann1998, duke2020, guan2012, qiii2015}. It provides efficient and robust solutions of the standard pairing Hamiltonian for both spherical and deformed systems. The key to its success is a procedure that can determine the initial guesses for the large set nonlinear equations involved in a controllable and physically motivated manner. In Table~\ref{hspa}, we present the ground-state eigenvalues from the PairDiagSph and Richardson-Gaudin method for systems with the constant pairing interaction strengths $G=0.1$ and pair numbers from 1 to 6. Within the given accuracy, the results from the two calculations are consistent.

\begin{table}
\small
\begin{center}
\setlength{\tabcolsep}{2.9mm}{
\caption{\label{diag} In the system of 7 pairs, numerical comparisons between PairDiagSph and PairDiag. $G$ are the constant pairing interaction strength, $E_{PairDiagS\!ph}$ and $E_{PairDiag}$ are the ground-state eigenvalues, $\Delta_{occup}$ is defined as $\sum|O_{PairDiagS\!ph}(i)-O_{PairDiag}(i)|$ where $O_{PairDiagS\!ph}$ and $O_{PairDiag}$ are the calculated occupation munbers of each orbital.}
\begin{tabular}{clll}
\hline
$G$             &  $E_{PairDiagS\!ph}$           &  $E_{PairDiag}$          &  $\Delta_{occup}$    \\
\hline
\texttt{-}0.2   &  \texttt{+}12.102028246        &  \texttt{+}12.102028247  &  1$\times$10$^{-6}$  \\
\texttt{-}0.4   &  \texttt{-}32.017674508        &  \texttt{-}32.017674507  &  4$\times$10$^{-5}$  \\
\texttt{-}0.6   &  \texttt{-}89.528347229        &  \texttt{-}89.528347233  &  3$\times$10$^{-6}$  \\
\hline
\end{tabular}}
\end{center}
\end{table}

For the pairing Hamiltonian in Eq.~(\ref{hamiltonian2}), if we only consider the time-reversal double degeneracy as the Hamiltonian in Eq.~(\ref{hamiltonian1}), the exact diagonalization can also be achieved in the Fock space, and the two different considerations should bring the same result for the same system. For the exact pairing solution in time-reversal-invariant systems, we developed a program, PairDiag~\cite{liuu2020}, to calculate the ground-state eigenvalue and the occupation numbers from the ground-state eigenvector. In Table~\ref{diag}, we present the ground-state eigenvalues and the difference of occupation numbers from the PairDiagSph and PairDiag for calculations in the system with 7 pairs under different constant pairing interaction strengths $G$. For this system, the dimension of the quasi-spin-space basis in PairDiagSph is 113372, while the dimension of the fermionic Fock-space basis used in PairDiag reaches 154143080. To compare the occupation numbers with PairDiagSph, we packaged the 53 occupancies from PairDiag into 16. As can be seen from Table~\ref{diag}, the results from the two methods are almost the same, but for spherical pairing Hamiltonian~\ref{hamiltonian2} PairDiagSph is more advantageous because of the quasi-spin symmetry considered.

\subsection{Running Time}

\begin{figure}
\begin{center}
\setlength{\abovecaptionskip}{0.0cm}
\includegraphics[width=0.44\textwidth]{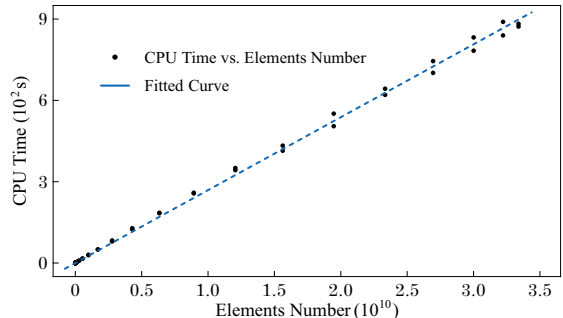}
\caption{\label{TvsE} The CPU time per Lanczos iteration with the hash search as a function of the total non-zero matrix elements number. The solid points are from the measurements and the dotted curve, $y=2.690\!\times\!10^{-8}x$, is the result of fitting.}
\end{center}
\end{figure}

The most time-consuming part of the calculation is the matrix-vector multiplication in Lanczos iterations. Therefore, the running time of the entire calculation mainly depends on the total number of iterations and the time cost per iteration. The total number of iterations can vary depending on interactions, spaces, and also the user's choice of error tolerance, usually around 50 iterations are needed for a good convergence of the ground state. The running time of a single iteration is expected to be proportional to the total number of non-zero elements in the Hamiltonian matrix in the use of hash search algorithm. To map the running time of the program, we performed 53 calculations corresponding to the pair numbers varies from 1 to 53 in the model space consisting of 16 orbitals from magic 20 to 126. Fig.~\ref{TvsE} represents the relationship between the CPU time per iteration and the total number of non-zero matrix elements from these calculations with the hash search used, in which the data shows a good linear relationship.

\begin{figure}
\begin{center}
\setlength{\abovecaptionskip}{0.0cm}
\includegraphics[width=0.44\textwidth]{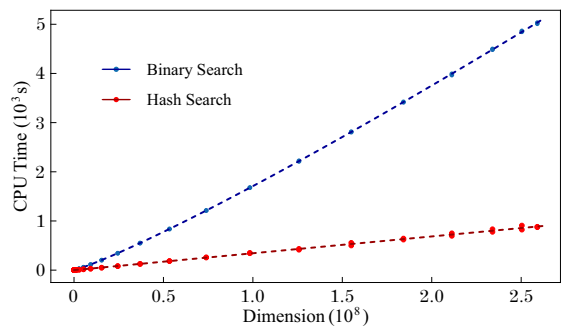}
\caption{\label{HvsB} The CPU time per Lanczos iteration as a function of the dimension with the hash search (red dots) and the binary search (blue dots). The red curve, $y=3.624\!\times\!10^{-6}x^{0.997}$, is the fitting for hash search, and the blue curve, $y=1.409\!\times\!10^{-6}x^{1.135}$, is the fitting for binary search.}
\end{center}
\end{figure}

For a sorted array, the binary search~\cite{liuu2020} can always be used to locate elements and this search algorithm has also been encoded into the PairDiagSph program for verification and comparison. Fig.~\ref{HvsB} shows a comparison of the CPU time per iteration between using the hash search and the binary search in the 53 calculations described above, from which we can see that the hash search has higher efficiency and better linearity. The CPU time is not the actual clock time, 1 iteration with dimension 2.6$\times\!10^{8}$ costs about 900 seconds of CPU time, but it actually only takes about 2 minutes in clock when eight CPU cores work in parallel. So, in the full calculation of this case (\texttt{Lanc\_\,Limit} is set to 20 and the memory cost is about 33GB), the total running time until having the ground state converged is about 1 hour and a half with 41 times of iterations in 2 restarts.

\section{Summary}

We presented an efficient diagonalization program for solving the general spherical pairing Hamiltonian based on the SU(2) quasi-spin algebra. Basis vectors with quasi-spin symmetry considered are generated by using the so-called adjacency excitation algorithm we developed. The Hamiltonian matrix constructed is diagonalized with the Lanczos + QR algorithm. All non-zero matrix elements for the matrix-vector multiplication are evaluated dynamically by the scattering operator and hash search actiong on the basis. With the OpenMp parallel Fortran module, PairDiagSph, developed by applying above algorithms, one can efficiently calculate the ground-state eigenvalue and eigenvector of the spherical pairing Hamiltonian for the system with fixed seniority. The total pair capacity of the program is 63 which meets the general needs for nuclear physics, and the calculation for spaces with dimension up to 10$^{8}$ can be done within hours on standard desktop computers.

\section{Acknowledgement}

The work was supported by the China Scholarship Council (201700260183) and the Liaoning Provincial Universities Overseas Training Program (2019GJWYB024).

\appendix

\section{A Simple Example of Using the PairDiagSph Module}\label{example}

A Fortran program for the spherical paring Hamiltonian using the PairDiagSph module (the model space in the calculation consists of 4 orbitals $1f_{7/2}$, $2p_{3/2}$, $1f_{5/2}$, and $2p_{1/2}$, giving the degeneracy $\Omega$ = \{4, 2, 3, 1\}).

\begin{tabular}{l}
\\
\textcolor[rgb]{0,0,1}{! Step 0: Declare a variable of the type Diag\_\,Par}                 \\
\textcolor[rgb]{0.7,0,0}{\textbf{use}} PairDiagSph                                           \\
\textcolor[rgb]{0.7,0,0}{\textbf{implicit none}}                                             \\
\textcolor[rgb]{0.7,0,0}{\textbf{type}}(Diag\_\,Par):: P1                                    \\
\end{tabular}

\begin{tabular}{l}
\textcolor[rgb]{0,0,1}{! Step 1: Initialize the input part}                                  \\
\textcolor[rgb]{0,0.5,0}{\textbf{integer}}(\textcolor[rgb]{0.7,0,0}{\textbf{kind}}=\textcolor[rgb]{1,0,0.7}{1}):: i, j \\
P1\%Shell = \textcolor[rgb]{1,0,0.7}{4}                                                      \\
P1\%Pairs = \textcolor[rgb]{1,0,0.7}{5}                                                      \\
P1\%Omega(\textcolor[rgb]{1,0,0.7}{1}:P1\%Shell) = (/\textcolor[rgb]{1,0,0.7}{4}, \textcolor[rgb]{1,0,0.7}{2}, \textcolor[rgb]{1,0,0.7}{3}, \textcolor[rgb]{1,0,0.7}{1}/)\\
\textcolor[rgb]{0.7,0,0}{\textbf{do}} i=\textcolor[rgb]{1,0,0.7}{1}, P1\%Shell               \\
\quad P1\%Senio(i) = \textcolor[rgb]{1,0,0.7}{0}                                             \\
\quad P1\%SPE(i) = i*\textcolor[rgb]{1,0,0.7}{1}                                             \\
\quad \textcolor[rgb]{0.7,0,0}{\textbf{do}} j=\textcolor[rgb]{1,0,0.7}{1}, P1\%Shell         \\
\qquad P1\%P\_\,F(i, j) = \textcolor[rgb]{1,0,0.7}{-0.2}                                     \\
\quad \textcolor[rgb]{0.7,0,0}{\textbf{end do}}                                              \\
\textcolor[rgb]{0.7,0,0}{\textbf{end do}}                                                    \\
\end{tabular}

\begin{tabular}{l}
\textcolor[rgb]{0,0,1}{! Step 2: Call the subroutine}                                        \\
\textcolor[rgb]{0.7,0,0}{\textbf{call}} Diag\_\,Solver(P1)                                   \\
\end{tabular}

\begin{tabular}{l}
\textcolor[rgb]{0,0,1}{! Step 3: Use the output part}                                        \\
\textcolor[rgb]{0.7,0,0}{\textbf{write}}(*, *) P1\%Energy\_\,Ground                          \\
\textcolor[rgb]{0.7,0,0}{\textbf{write}}(*, *) P1\%Monopole\_\,Min                           \\
\textcolor[rgb]{0.7,0,0}{\textbf{do}} i=\textcolor[rgb]{1,0,0.7}{1}, P1\%Shell               \\
\quad \textcolor[rgb]{0.7,0,0}{\textbf{write}}(*, *) P1\%N\_\,Occup(i)                       \\
\textcolor[rgb]{0.7,0,0}{\textbf{end do}}                                                    \\
\end{tabular}

\section{Brief Description of Variables and Subroutines}\label{brief}

Variables:

\begin{itemize}
\setlength{\itemsep}{0pt}
\setlength{\parsep}{0pt}
\setlength{\parskip}{0pt}
\item Lanc\_\,Limit: The size of the Lanczos iteration subspace.
\item Lanc\_\,Error: In restart mode, the convergence threshold in $|\,\beta_{i}/\alpha_{i}| \le$ Lanc\_\,Error.
\item N\_\,Total, N\_\,Shell, and N\_\,Pairs: The total number of orbitals, shells, and pairs in the calculation.
\item B\_\,Dimension and L\_\,Dimension: The Dimension of the basis space and the Lanczos iteration subspace in the calculation.
\item Convergence: Flags for convergence for the restart mode.
\item Run\_\,Mode and Print\_\,Mode: Flags for run and print.
\item Energy\_\,Senio: The energy from unpaired particles corresponding to the second term in Eq.~(\ref{diagonal}). 
\item Energy\_\,Ground: The output ground-state eigenvalue.
\item Monopole\_\,Min: The minimum of the diagonal elements.
\item Posit\_\,Min: The position of the vector with the minimum diagonal element.
\end{itemize}

Arrays:

\begin{itemize}
\setlength{\itemsep}{0pt}
\setlength{\parsep}{0pt}
\setlength{\parskip}{0pt}
\item SPE: The 1D array for single-particle energies.
\item P\_\,F: The 2D array for pairing strength.
\item N\_\,Omega: The 1D array for the degeneracy. 
\item P\_\,Index: The 1D array for the indexes.  
\item B\_\,Array: The 1D array for the basis vectors. 
\item C\_\,Array: The 2D array for coefficients $N_{p}^{d}$ for the hash search. 
\item Q\_\,Matrix: The 2D array for the Lanczos/Ritz vectors. 
\item L\_\,Matrix: The 2D array for the Lanczos Matrix and eigenvalues. 
\item N\_\,Occup: The 1D array for occupation numbers.
\item I\_\,Vector and Q\_\,Vector: The temporary 1D arrays for Lanczos iteration. 
\item O\_\,Array, V\_\,Array, O\_\,Prray, and V\_\,Prray: The temporary 1D arrays for vector search.  
\item R\_\,Array: The temporary 1D array for vector search.  
\item T\_\,Matrix and P\_\,Matrix: The temporary 2D arrays for QR decompositon. 
\item Omega\_\,I and Omega\_\,F: The 1D arrays for the positions of the first and the last digits in shells.  
\item E\_\,Array: The 1D array for the frist 10 eigenvalues.  
\end{itemize}

Type, subroutines, and functions:

\begin{itemize}
\setlength{\itemsep}{0pt}
\setlength{\parsep}{0pt}
\setlength{\parskip}{0pt}
\item Diag\_\,Par: Derived data type.
\item Diag\_\,Solver(Diag\_\,Par, [Mode]): The public subroutine that starts the calculation.
\item Initialize(): The subroutine that allocates memory for dynamic arrays and initializes basis vectors.
\item Monopole\_\,E(State): The function returns the diagonal element value of input (State) according to Eq.~(\ref{diagonal}).
\item Next\_\,State(State): The subroutine operates the input (State) according to the Adjacency excitation algorithm.~\ref{conversion}.
\item Bina\_\,State(D, L) and Hash\_\,State(D, L): The subroutines that calculate non-zero matrix elements and positions related to the input State using binary and hash search.
\item Vector\_\,Initialize() and Vector\_\,Restart(): The subroutines that initialize the starting vector to $[1,0,\cdots,0]^{T}$ and Q\_\,Matrix(1, :).
\item Lanczos\_\,Iteration(): The subroutine for Lanczos iteration from starting vector I\_\,Vector.
\item QR\_\,Decompose(): The subroutine for QR decompose to the L\_\,Matrix.
\item Destory(): The subroutine that releases all dynamic memories.
\item Lanczos\_\,QR(): The subroutine that combines the Lanczos\_\,Iteration() and QR\_\,Decompose().
\item Results\_\,Output(): The subroutine that calculates all the outputs.
\end{itemize}

\end{document}